\documentclass[doublecol]{epl2} 

\newcommand{\commentout}[1]{}
\commentout{
\documentclass[11pt]{amsart}

\newcommand{\commentout}[1]{}
\reversemarginpar
\pagestyle{plain}
\setlength{\textwidth}{16.5truecm}
\setlength{\textheight}{22.5truecm}
\setlength{\topmargin}{-0.5truecm}
\setlength{\oddsidemargin}{0cm}
\setlength{\evensidemargin}{\oddsidemargin}

}

\usepackage{amsmath}
\usepackage{amssymb}

\newcommand{\nwc}{\newcommand}

\newcommand{\lt}{\left}
\nwc{\partz}{\frac{\partial }{\partial z}}
\newcommand{\rt}{\right}
\nwc{\ytil}{\tilde{\by}}
\nwc{\al}{\alpha}
\nwc{\half}{\frac{1}{2}}

\newcommand{\lan}{\left\langle}
\newcommand{\ran}{\right\rangle}

\newcommand{\ks}{{k}}
\newcommand{\bx}{\mathbf x}
\newcommand{\bD}{\mathbf D}
\newcommand{\bp}{\mathbf p}
\newcommand{\br}{\mathbf r}
\newcommand{\bq}{\mathbf q}
\newcommand{\by}{\mathbf y}

\nwc{\nwt}{\newtheorem}

\nwt{proposition}{Proposition}
\nwt{lemma}{Lemma}
\nwt{theorem}{Theorem}

\nwt{remark}{Remark}
\nwt{assumption}{Assumption}
\nwt{definition}{Definition} 

\nwc{\bal}{\begin{align}}
\nwc{\be}{\begin{equation}}
\nwc{\ben}{\begin{equation*}}
\nwc{\bea}{\begin{eqnarray}}
\nwc{\beq}{\begin{eqnarray}}
\nwc{\bean}{\begin{eqnarray*}}
\nwc{\beqn}{\begin{eqnarray*}}
\nwc{\beqast}{\begin{eqnarray*}}

\nwc{\eal}{\end{align}}
\nwc{\ee}{\end{equation}}
\nwc{\een}{\end{equation*}}
\nwc{\eea}{\end{eqnarray}}
\nwc{\eeq}{\end{eqnarray}}
\nwc{\eean}{\end{eqnarray*}}
\nwc{\eeqn}{\end{eqnarray*}}
\nwc{\eeqast}{\end{eqnarray*}}

\nwc{\invf}{\cF^{-1}_2}
\nwc{\ep}{\varepsilon}
\nwc{\tep}{\tilde{\varepsilon}}
\nwc{\epsq}{{\varepsilon^2}}
\nwc{\epsqa}{{\varepsilon^{2\alpha}}}
\nwc{\eps}{\varepsilon}
\nwc{\ept}{\epsilon}
\nwc{\vrho}{\varrho}
\nwc{\orho}{\bar\varrho}
\nwc{\ou}{\bar u}
\nwc{\vpsi}{\varpsi}
\nwc{\lamb}{\lambda}

\nwc{\nn}{\nonumber}
\nwc{\mf}{\mathbf}
\nwc{\mb}{\mathbf}
\nwc{\ml}{\mathcal}

\nwc{\IA}{\mathbb{A}} 
\nwc{\IB}{\mathbb{B}}
\nwc{\IC}{\mathbb{C}} 
\nwc{\ID}{\mathbb{D}} 
\nwc{\IM}{\mathbb{M}} 
\nwc{\IP}{\mathbb{P}} 
\nwc{\II}{\mathbb{I}} 
\nwc{\IE}{\mathbb{E}} 
\nwc{\IF}{\mathbb{F}} 
\nwc{\IG}{\mathbb{G}} 
\nwc{\IN}{\mathbb{N}} 
\nwc{\IQ}{\mathbb{Q}} 
\nwc{\IR}{\mathbb{R}} 
\nwc{\IT}{\mathbb{T}} 
\nwc{\IZ}{\mathbb{Z}} 

\nwc{\epal}{\ep^{-2\alpha}}
\nwc{\cE}{{\ml E}}
\nwc{\cP}{{\ml P}}
\nwc{\cQ}{{\ml Q}}
\nwc{\cL}{{\ml L}}
\nwc{\cR}{{\ml R}}
\nwc{\cV}{{\ml V}}
\nwc{\cT}{{\ml T}}
\nwc{\crV}{{\ml L}_{(\delta,\rho)}}
\nwc{\cC}{{\ml C}}
\nwc{\cA}{{\ml A}}
\nwc{\cK}{{\ml K}}
\nwc{\cB}{{\ml B}}
\nwc{\cD}{{\ml D}}
\nwc{\cF}{{\ml F}}
\nwc{\cS}{{\ml S}}
\nwc{\cW}{{\ml W}}
\nwc{\cG}{{\ml G}}
\nwc{\cH}{{\ml H}}
\nwc{\bk}{{\mb k}}
\nwc{\cbz}{\overline{\cB}_z}
\nwc{\fW}{\mathfrak{W}}

\nwc{\fF}{\mathfrak{F}}
\nwc{\pft}{\cF^{-1}_\bp}

\title{Space-frequency correlation of classical waves in disordered media: high-frequency and small scale asymptotics}
\shorttitle{Space-frequency  correlations} 

\author{A. C. Fannjiang\inst{1}}
\shortauthor{A. C. Fannjiang}

\institute{                    
  \inst{1} Department of Mathematics,
University of California,
Davis, CA 95616-8633
}
\pacs{42.25.Dd}{Wave propagation in random media}
\pacs{42.68.Ay}{Propagation, transmission, attenuation, and radiative transfer}
\pacs{05.40.-a}{Fluctuation phenomena, random processes, noise, and Brownian motion}

\abstract{Two-frequency radiative transfer (2f-RT) theory is developed for geometrical optics in random media. The space-frequency correlation is described by the two-frequency Wigner distribution (2f-WD) which satisfies a closed form
equation, the two-frequency Wigner-Moyal equation. 
In the RT regime it is proved rigorously 
that 2f-WD satisfies
a Fokker-Planck-like equation with complex-valued coefficients. 
By dimensional analysis 2f-RT equation yields
the scaling behavior of 
three physical parameters: the spatial spread,
the coherence length and the coherence bandwidth. 
The sub-transport-mean-free-path behavior is obtained in a closed form by analytically solving
a paraxial 2f-RT equation. 
}

\begin{document}

\maketitle

\section{Introduction}

Correlation functions of fields arise
naturally in the description of fluctuations and  are 
ubiquitous objects in statistical physics. 
The most basic of those are the second-order correlations
in the space-time or space-frequency domain; the two
are equivalent to each other via the Fourier transform.  When
the field fluctuations can be described as a Gaussian
stochastic process, all the correlation functions
of the field can then be expressed in term of
the second order ones, by the use of
the moment theorem for Gaussian processes. 
The second order space-frequency correlation
then emerges as an  indispensable tool for studying 
fluctuations of fields and
is equivalent to the {\em mutual coherence function}
describing the field correlation at
two space-time  points \cite{MW}.

Spatial and temporal structures of ultrawide-band high-frequency fields can be appreciably affected by small random changes of the medium parameters characteristic of almost all astro-
and geophysical environments. An  important step
toward analytical understanding of pulse propagation in multiply scattering media
is then to derive the equation for the space-frequency
correlation, obtain the qualitative information about
its behavior and, if possible,  find its (asymptotic) solutions. This problem has been extensively
studied in the literature, see, e.g., \cite{Ish, BF, RN, SF, BM}.
The main distinction of our approach from previous works is that 
our approach to space-frequency correlation  is carried out in terms of the two-frequency
Wigner distribution (2f-WD)
for which we will
derive rigorously  equations of relatively simple form in the  radiative transfer (RT) regime  and
obtain an exact solution for the small-scale behavior below the transport mean-free-path \cite{MW,Cha}.  

The standard (equal-time or -frequency) Wigner
distribution (WD) is a quasi-probability density function in phase space and was first introduced by E. Wigner \cite{Wig} 
in connection to quantum thermodynamics and later found
wide-ranging applications in classical \cite{Dra}, \cite{josa},  as
well as in quantum optics \cite{MW}, \cite{Sch}. In classical optics,
a main use
of the Wigner distribution is connected  to
high-frequency asymptotic and radiative transfer, both of which
can be most naturally worked out from the first principle  in phase space (see the review  \cite{KA}, \cite{rad-trans} and references therein). 


The main advantage of 2f-RT over the traditional equal-time radiative transfer theory is that it describes  not just the energetic transport  but also the two space-time point mutual coherence in the following way. 

Let the scalar wave field $U_j, j=1,2,$ of 
wavenumber
$k_j, j=1,2.$ 
be governed  by the reduced wave equation
\beq
\label{helm}
\Delta U_j(\br)+k_j^2 \big(\nu+ V(\br)\big)U_j(\br)=0, \quad\br\in \IR^3, \quad j=1,2
\eeq
where $\nu$ and $V$ are respectively  the mean 
and fluctuation of the refractive index associated
are assumed to be real-valued, corresponding to a lossless medium. 
For simplicity,
we restrict our attention to dispersionless media
(see \cite{2f-rt-josa} for discussion on
dispersive media). 
Here and below the background wave speed is set to be
 unity so that $k_{j}=\omega_{j}$. 
Let $u(t_j,\bx_j), j=1,2$ be the time-dependent wave field
at two space-time points $(t_j, \bx_j), j=1,2.$ Let
$\bx=(\omega_1\bx_1+\omega_2\bx_2)/2$ and $\by=\omega_1\bx_1-\omega_2\bx_2$. Then we have
\beq
\label{2time}
&&\lan u(t_1,\bx_1) u^*(t_2,\bx_2)\ran\\
&=& \int e^{i\bp\cdot\by}  e^{i(\omega_2t_2-\omega_1t_1)} \lan W(\bx,\bp; \omega_1, \omega_2)\ran
 d\omega_1 d\omega_2 d\bp\nn
\eeq
where $W(\bx,\bp; \omega_1, \omega_2)$ is  the 2f-WD  defined by 
\beqn
\lefteqn{W(\bx,\bp;\omega_{1},\omega_{2})}\\
&=&\frac{1}{(2\pi)^3}\int
e^{-i\bp\cdot\by}
U_1 (\frac{\bx}{\omega_1}+
\frac{\by}{2\omega_1}){U^*_2(\frac{\bx}{\omega_2}
-\frac{\by}{2\omega_2})}d\by\nn\\
&=&\big(\omega_1\omega_2\big)^3\int e^{i\bx\cdot\bq}\hat U_1\Big(\omega_1\bp+\frac{\omega_1\bq}{2}\Big)
\hat U_2^*\Big(\omega_2\bp-\frac{\omega_2\bq}{2}\Big)
d\bq.\nn
\eeqn
Here and below $\lan\cdot\ran$ denotes the ensemble average. 
For {\em temporally stationary}
signals, wave fields of different frequencies are uncorrelated
and only the equal-frequency WD is necessary to describe the two-time correlation. 
 In comparison, the single-time correlations
with $t_1=t_2=t$ gives rise to the expression 
\beqn
&&\lan u(t,\bx_1) u^*(t,\bx_2)\ran
= \int  d\omega'd\bp \,\, e^{i\bp\cdot\by}e^{i\omega' t}\\
&&\times\int d\omega \lan W(\bx,\bp; \omega-\omega'/2, \omega+\omega'/2)\ran 
\eeqn
which is equivalent to the central-frequency-integrated  2f-WD.
For ease of notation, we will drop the frequency arguments
when writing the 2f-WD below.

\section{Weak coupling limit}

The radiative transfer regime sets in when the scale of medium fluctuation  is much smaller than
the propagation distance but is comparable or
much larger than the  wavelength. Based on
the general principle of central limit theorem,
RT corresponds to  the scaling limit
which replaces $\nu+V$ in eq. (\ref{helm}) with
\beq
\label{scaling}
\frac{1}{\theta^2\ep^2}\Big(\nu+ \sqrt{\ep}V(\frac{\br}{\ep})
\Big),\quad \theta>0,\quad\ep\ll 1
\eeq
 where $\ep>0$ and $\theta^{-1}>0$ are, respectively,
the ratio of the scale of medium fluctuation to the propagation distance and 
 the wavelength. Thus, $\ep\theta$ is the ratio of the wavelength to the
 propagation distance and as a result we rescale 
  the wavenumber as 
$k\to k/(\ep\theta)$, giving rise to 
the prefactor $(\theta\ep)^{-2}$. 
This is so called the weak coupling (or disorder) limit in kinetic theory
\cite{Spo} under which the Anderson localization
can not take place.  

We assume that $V(\bx)$ is  an ergodic, mean-zero,
statistically homogeneous random field. As a consequence,
$V$  admits the
spectral representation $V(\bx)=\int e^{{i\bx\cdot\bp}}
\hat{V}(d\bp)$ where   the spectral measure $\hat V$ satisfies
$
\lan \hat{V}(d\bp)\hat{V}(d\bq)\ran
=\delta(\bp+\bq)\Phi(\bp)d\bp d\bq
$
with $\Phi$ is the power spectral density. 
Since $V$ is real-valued,
 $\Phi(\bp)$
is real-valued,  non-negative and possesses the symmetry
 $\Phi(\bp)=\Phi(-\bp),\forall \bp$.

Physically  speaking radiative transfer  belongs to   the diffusive wave  regime under the condition of a large dimensionless conductance $g\gg 1$.
Let $A$ be   the illuminated area, $\lambda$ the wavelength of
radiation 
and $L$ the distance of propagation.
Let $N_f=\lambda L/A$ be the Fresnel number
and $\ell_*$  the transport mean-free path. 
The dimensionless conductance can then be expressed
simply as $g=k\ell_* /N_f $.
With the scaling (\ref{scaling}), $k\ell_* \sim N_f^{-1}\sim \theta^{-1}\ep^{-1}$
 and hence
$g\sim \theta^{-2}\ep^{-2}\gg 1$ for $\theta\ep\ll 1$.

To adapt to the weak coupling and the geometrical optics (see below) scalings  we introduce the two parameters $\ep, \theta$ into 
the 2f-WD and redefine it as
\beq
\label{2fw}
\lefteqn{W^\ep(\bx,\bp)}\\
&=&\frac{1}{(2\pi)^3}\int
e^{-i\bp\cdot\by}
U_1 (\frac{\bx}{k_1}+
\frac{\theta\ep\by}{2k_1}){U^*_2(\frac{\bx}{k_2}
-\frac{\theta\ep\by}{2k_2})}d\by\nn
\eeq
 In view of  the definition, we see
 that both $\bx$ and $\bp$ are dimensionless.  
 The particular scaling factors are introduced in (\ref{2fw})
 so that $W^\ep$ satisfies the following Wigner-Moyal equation
{\em  exactly} \cite{2f-rt-josa}
\beq
\label{wig}
{\bp\cdot\nabla W^\ep}
&=&\frac{1}{\sqrt{\ep}}\cL W^\ep
\eeq
where the operator $\cL$ is defined by
\beqn
\cL W^\ep(\bx,\bp)&=&
\frac{i}{2\theta}\int \hat V(d\bq)
e^{i\frac{\bq\cdot\bx}{\ep k_1}} W^\ep(\bx,\bp-\frac{\theta\bq}{2k_1})\\
&&-\frac{i}{2\theta}\int \hat V(d\bq)
e^{i\frac{\bq\cdot\bx}{\ep k_2}} W^\ep(\bx,\bp+\frac{\theta\bq}{2k_2}).\nn
\eeqn
In contrast, the Sudarshan equations for the mutual coherence function are, like (\ref{wig}),  first-order in time but nonlocal
in space even in the case of  free field \cite{MW}.

\section{High-frequency regime}
Before we consider  the radiative transfer limit $\ep\downarrow 0$
further let us take the high-frequency limit  $\theta \downarrow 0$
while maintaining the following relationships 
\beq
\nn
\lim_{\theta\to 0}\ks_1=\lim_{\theta\to 0}\ks_2&=&\ks\\
\frac{\ks_2-\ks_1}{\theta\ep k}&=&\beta \label{range}\label{band2}
\eeq
where $\beta>0$ is independent of $\theta$ and $\ep$, representing the normalized difference  in wavenumber.  Frequencies within the range described by (\ref{range})   remain coherent
with one another.

 In this regime, we see from (\ref{2fw}) that to leading order
the center of two field points is $\bx/k$ and the difference
is $\theta\ep (\by+\beta\bx)/k$. 
Passing to the limit $\theta\downarrow 0$ in (\ref{wig}) we obtain
the first-order partial differential equation
\beq
\label{wig-go}
\lefteqn{{\bp\cdot\nabla_\bx W^\ep}(\bx,\bp)}\\
&=&-\frac{1}{2k\sqrt{\ep}}\big(\nabla V\big)\Big(\frac{\bx}{\ep k}\Big)\cdot\Big[\nabla_\bp 
-i\beta  \bx \Big]W^\ep(\bx,\bp).\nn
 \eeq
 For $\beta=0$, eq. (\ref{wig-go}) is the static Liouville
equation. For $\beta>0$, eq. (\ref{wig-go}) retains
the wave character and  is the focus of the subsequent 
analysis. We shall refer to eq. (\ref{wig-go}) as
the two-frequency Liouville equation (2f-LE). 

Consider, for instance, the WKB ansatz
\beqn
U_j(\br)=A_j(\br) \exp{\Big(\frac{ik_j}{\theta\ep}S_j(\br)\Big)},\quad j=1,2
\eeqn
where the phase $S_j$ and the amplitude
$A_j$ depend on the frequency differentiably.  
In the first case, assume $S_1=S_2=S$. Then  in the high-frequency limit
2f-WD becomes
\beq
\label{go-data}
W^\ep(\bx,\bp)
= e^{i\beta \bx\cdot\bp}e^{-i\beta k S(\frac{\bx}{k})}
|A|^2 \Big(\frac{\bx}{k}\Big) \delta\Big(\bp-\nabla S\big(\frac{\bx}{k}\big)\Big)
\eeq
which satisfies  2f-LE. In the second case, assume $S_j(\br)=\hat\bk_j\cdot\br, |\hat\bk_j|=1$, with
the additional conditions 
\beq
\label{plane1}
\lim_{\theta\to 0}\hat\bk_1&=&
\lim_{\theta\to 0}\hat\bk_2=\hat\bk\\
\frac{\hat\bk_2-\hat\bk_1}{\theta\ep}&=& \Delta\hat\bk
\label{plane2}
\eeq
where $\Delta\hat\bk$ is independent of $\theta,\ep$.
 Then the 
the 2f-WD becomes
\beq
\label{plane3}
|A|^2\Big(\frac{\bx}{k}\Big)e^{i\Delta\hat\bk\cdot \bx}\delta(\bp-\hat\bk)
\eeq
where $\beta$ is absent due to the linear phase  profile  $S_j$. 

Given, say,  the Dirichlet boundary condition $F$ imposed on
the boundary $\partial \cD$ of a phase-space domain $\cD$,  2f-LE can be
solved by the method of characteristics as shown below.
The form of 2f-LE suggests the ``gauge transformation''
of 2f-WD 
 \beq
 \label{transf}
 \fW^\ep(\bx,\bp)=e^{-i\beta\bx\cdot\bp}W^\ep(\bx,\bp)
 \eeq
 which then satisfies the following more convenient equaiton
  \beq
\lefteqn{\bp\cdot\nabla_\bx \fW^\ep+i\beta|\bp|^2 \fW^\ep}\label{fwe}\\
&=&-\frac{1}{2k\sqrt{\ep}}\big(\nabla V\big)\Big(\frac{\bx}{\ep k}\Big)\cdot\nabla_\bp \fW^\ep\nn
 \eeq
with the boundary condition that $\fW^\ep(\bx,\bp)=\exp{[-i\beta\bx\cdot\bp]}F(\bx,\bp)\equiv\fF(\bx,\bp)$ on $\partial \cD$. 
In view of (\ref{transf}) $\fW^\ep$ is the Fourier transform
of the two-point function $U_1\otimes U_2^*$
in the location difference (i.e. $\by+\beta\bx$ measured in the unit of the
central wavelength).

  Consider
the  Hamiltonian system of time-reversed characteristic curves
\beq
\frac{d}{dt} \bx^\ep(t)&=&-{\bp^\ep(t)}\label{ham1}\\
\frac{d}{dt} \bp^\ep(t)&=&-\frac{1}{2k\sqrt{\ep}}\big(\nabla V\big)\Big(\frac{\bx^\ep(t)}{k\ep}\Big)\label{ham2}
\eeq
with $\bx^\ep(0)=\bx,\bp^\ep(0)=\bp$.
Let $\tau^\ep=\tau^\ep(\bx,\bp)$ be the first passage time when
the trajectory $(\bx^\ep(\cdot), \bp^\ep(\cdot))$ 
hits the boundary of the phase-space domain  $\cD$. 
We then have
\beqn
\lefteqn{\fW^\ep(\bx,\bp)}\\
&=&e^{-i\beta \int^{\tau^\ep}_0|\bp^\ep(s)|^2ds
-i\beta\bx^\ep(\tau^\ep)\cdot\bp^\ep(\tau^\ep)}F(\bx^\ep(\tau^\ep),\bp^\ep(\tau^\ep)).
\eeqn

\section{Convergence to diffusion in momentum}
If  $V$ decorrelates sufficiently rapidly
(see \cite{KP} for  a precise formulation), the probability distribution of $(\bx^\ep(\cdot),\bp^\ep(\cdot))$
 defined by (\ref{ham1})-(\ref{ham2}),
converges weakly, as $\ep\to 0$, to that of
the Markov process $(\bx(\cdot),\bp(\cdot))$ where
\beq
\label{disp}
\bx(t)=\bx-\int^t_0\bp(s)ds
\eeq
and $\bp(\cdot), \bp(0)=\bp,$ is a diffusion process generated by the operator
\beqn
\cA=\frac{1}{4k}\nabla_\bp\cdot \bD\cdot\nabla_\bp
\eeqn
with the (momentum) diffusion coefficient
\beq
\label{diff-coeff}
\bD(\bp)=\pi\int \Phi(\bq)\delta(\bp\cdot\bq)\bq\otimes\bq d\bq.
\eeq
Writing $\bD$  as 
\beq
\label{diff-coeff2}
\bD(\bp)=\pi\int \Phi(\bq)\delta(\bp\cdot\bq)\Pi(\bp)\bq\otimes \Pi(\bp)\bq d\bq
\eeq
where $\Pi(\bp)$ is the orthogonal projection onto
the hyperplane perpendicular to $\bp$ we see
that 
 the momentum diffusion process is concentrated
on the sphere of radius $|\bp|$. In other words, the limiting
kinetic energy $|\bp(t)|^{2}/2$ is preserved by  the elastic scattering process. This observation will be  useful
for the subsequent calculation.

The consequence is the convergence of the ensemble average
$\lan \fW^\ep(\bx,\bp)\ran $ to 
\beq
\label{eq19}
\lefteqn{\fW(\bx,\bp)}\\
&\equiv &\IE_{\bx,\bp}\Big\{e^{-i\beta |\bp|^2\tau-i\beta\bx(\tau)\cdot\bp(\tau)}F(\bx(\tau),\bp(\tau))\Big\}\nn
\eeq
where $\tau=\tau(\bx,\bp)$ is the first passage time of the Markov process
$(\bx(t),\bp(t))$ with $\bx(0)=\bx,\bp(0)=\bp$ and
 $\IE_{\bx,\bp}$ the corresponding average. 
 
Now let $W(\bx,\bp)$ be the solution of
the following boundary value problem:
 \beq
 \label{go} \label{fp}
{\bp\cdot\nabla_\bx W}
=\frac{1}{4k}\Big(\nabla_\bp-i\beta\bx\Big)\cdot\bD\cdot
\Big(\nabla_\bp-i\beta \bx\Big)W
\eeq
$ \hbox{with}\quad W=F\quad \hbox{on}\quad
\partial \cD $ and we will show
that the solution of 2f-RT is the pointwise limit
of the average 2f-WD.  Eq. (\ref{go}) is  our two-frequency
radiative transfer  (2f-RT) equation. 
Because we have considered the  high frequency asymptotics
the scattering term takes the form of a second-order differential operator rather than the more familiar integral operator.

Let $\bp(t)$ be the diffusion process generated by
the generator $\cA$  
and define
\beq
\label{go2}
\widetilde W(t, \bx,\bp) =\exp{\Big[-i\beta t|\bp|^2-i\beta\bx\cdot\bp\Big]}W(\bx,\bp). 
\eeq
By 
Dynkin's formula  \cite{SV} we have that
\beqn
&&\IE_{\bx,\bp}\Big\{\widetilde W(\tau,\bx(\tau),\bp(\tau))\Big\}=
\widetilde W(0,\bx,\bp)\\
&&+\IE_{\bx,\bp}\Big\{\int^\tau_0
\big[\frac{\partial}{\partial s}-\bp\cdot\nabla_\bx+\cA\big]\widetilde W(s, \bx(s),
\bp(s))ds\Big\}.\nn
\eeqn
From   (\ref{go})-(\ref{go2}) it follows that
\beqn
\big[\frac{\partial}{\partial t}-\bp\cdot\nabla_\bx+\cA\big]\widetilde W=0
\eeqn
and 
\beq
\label{ww}
\fW(\bx,\bp)=\IE_{\bx,\bp}\Big\{\widetilde W(\tau,\bx(\tau),\bp(\tau))\Big\}
=e^{-i\beta\bx\cdot\bp}
W(\bx,\bp). 
\eeq
 Therefore,  in view of (\ref{transf}),  
$W(\bx,\bp)$ is the pointwise limit of
$\lan W^\ep(\bx,\bp)\ran $. 
It is straightforward  to check that $\fW$ is the solution to the
equation
\beq
\label{pract}
\bp\cdot\nabla_\bx \fW+i\beta|\bp|^2 \fW
=\cA \fW.
\eeq

From (\ref{eq19}) and (\ref{ww}) we obtain 
\commentout{
\beqn
i\beta \int^t_0|\bp(s)|^2ds-i\beta\bx(t)\cdot\bp(t)
&=&i\beta \int^t_0\bp(s)d\bx(s)-i\beta\bx(t)\cdot\bp(t)\\
&=&-i\beta \int^t_0 \bx(s)d\bp(s)-i\beta\bx(0)\cdot\bp(0)
\eeqn
and hence by (\ref{eq19}) and (\ref{eq23})  the limit 
}  the probabilistic representation for $W$
\beq
\label{path}
W(\bx,\bp)=\IE_{\bx,\bp}\Big\{e^{-i\beta \int^\tau_0\bx(s)d\bp(s)}F(\bx(\tau),\bp(\tau))\Big\}
\eeq
where $\int^\tau_0\bx(s)d\bp(s)$ is  understood as an It\^o integral \cite{SV}. Expression  (\ref{path}) suggests 
a numerical solution procedure  for 2f-RT by Monte Carlo simulation.

\section{Isotropic medium}
Eq. (\ref{pract}) clearly is translationally invariant in $\bx$ due
to the stationarity of the medium. 
If the medium is also statistically isotropic, then
eq. (\ref{pract}) is rotationally invariant.
 To see this  let us consider an isotropic 
spectral density $\Phi(\bp)=\Phi(|\bp|)$. Then we have
 $\bD={C} |\bp|^{-1} \Pi(\bp)$ where 
 \beq
 \label{const}
{C}=\frac{\pi}{3}\int\delta\Big(\frac{\bp}{|\bp|}\cdot\frac{\bq}{|\bq|}\Big)\Phi(|\bq|)|\bq| d\bq
\eeq
is a constant. The coefficient $C$ (and $\bD$)  has the dimension of inverse length while the variables $\bx$ and $\bp$ are dimensionless. 

The resulting $\cA$ is invariant with respect to rotation in $\bp$.  Hence if $\fW(\bx,\bp)$ is a solution to (\ref{pract})  then 
$\fW(R \bx,R\bp)$ is also a solution  where
$R$ is any orthogonal matrix.

\section{Spatial (frequency) spread and coherence bandwidth}
Through dimensional analysis, eq. (\ref{go}) 
yields qualitative information about 
important physical parameters of the disordered medium.
For this, let us  assume for simplicity  the isotropy of the medium
as above. 

Now consider the following change of variables
\beq
\label{change}
\bx=\sigma_x k\tilde \bx,\quad \bp=\sigma_p\tilde\bp/k,
\quad \beta=\beta_c\tilde\beta
\eeq
where $\sigma_x$ and $\sigma_p$ are respectively the position
spread  and the spatial frequency spread, and $\beta_c$ is
the coherence bandwidth, also known as the Thouless frequency. Let us substitute (\ref{change})
into eq. (\ref{fp}) and aim for the normalized form
\beq
\label{stan}
{\tilde\bp\cdot\nabla_{\tilde\bx}W}=
\lt(\nabla_{\tilde\bp}-i{\tilde\beta}\tilde\bx\rt)
\cdot \frac{\Pi(\tilde\bp)}{|\tilde\bp|}
\lt(\nabla_{\tilde\bp}-i{\tilde\beta}\tilde\bx\rt)W.
\eeq
The 1-st term on the left side yields the first duality relation
$
\sigma_x/\sigma_p\sim 1/k^2.
$
The balance of the terms in each pair of the parentheses yields
the second duality relation
$
\sigma_x\sigma_p\sim 1/{\beta_c}
$
whose left hand side is the {\em space-spread-bandwidth product.}
Finally the removal of the constant $C$ determines
$\sigma_{p}$
from which 
 $\sigma_x$ and $\beta_c$ can be determined
 by using the duality relations. We obtain
 \beq
 \label{scaling}
\sigma_p\sim k^{2/3} C^{1/3}, \sigma_x\sim k^{-4/3} C^{1/3},\beta_c\sim k^{2/3} C^{-2/3}.
 \eeq
 
\section{Spatially anisotropic media}
Forward-scattering approximation, also called paraxial approximation,  is valid when  back-scattering is negligible
and, as we show below, this is the case for spatially anisotropic media fluctuating 
slowly  in the (longitudinal) direction of propagation. 

 Let  $z$ denote the longitudinal coordinate and $\bx_\perp$ the transverse coordinates. Let $p$ and $\bp_\perp$ denote the longitudinal and
transverse components of $\bp\in \IR^3$, respectively. 
Let  $\bq=(q, \bq_\perp)\in \IR^3$ be likewise defined. 
Consider now  a highly anisotropic spectral density
for  a medium fluctuating much more
slowly in the longitudinal direction, i.e.
replacing $\Phi\big(\bq\big)$ in (\ref{diff-coeff}) by 
$
{\eta^{-1}}\Phi\lt(\eta^{-1}q,  \bq_\perp\rt),\,\,\eta\ll 1,
$
which, in the limit $\eta\to 0$, tends to
\beq
\label{aniso}
\delta(q) \int dw \Phi\lt(w, 
\bq_\perp\rt). 
\eeq
We then obtain the transverse diffusion coefficient 
\beqn
\bD_\perp(\bp_\perp)=\pi \int d\bq_\perp \int dw \Phi(w,\bq_\perp)
\delta(\bp_\perp\cdot\bq_\perp)\bq_\perp\otimes\bq_\perp
\eeqn
whereas the longitudinal diffusion coefficient now vanishes. 
In other words, the longitudinal momentum is decoupled
from the transverse momentum and is not directly affected by
the multiple scattering process. 

For simplicity we  
 assume 
 the transverse isotropy, i.e. $\Phi(w,\bp_\perp)=\Phi(w,|\bp_\perp|)$, so that
 $\bD_\perp={C_\perp} |\bp_\perp|^{-1} \Pi_\perp(\bp_\perp)$ where 
 \beqn
{C_\perp}=\frac{\pi}{2}\int\delta\Big(\frac{\bp_\perp}{|\bp_\perp|}\cdot\frac{\bq_\perp}{|\bq_\perp|}\Big)\Phi(w, |\bq_\perp|)|\bq_\perp| dw d\bq_\perp
\eeqn
is a constant
and 
$\Pi_\perp(\bp_\perp)$ is 
the orthogonal projection onto the line perpendicular to
$\bp_\perp$.
Hence eq. (\ref{go}) reduces to
\beq
\label{go-para}
\lefteqn{\Big[p\partial_{z}+{\bp_\perp\cdot\nabla_{\bx_\perp} \Big]\bar W}}\\
&=&
\frac{C_\perp}{4k}\lt(\nabla_{\bp_\perp}-i{\beta}\bx_\perp\rt)\cdot \frac{\Pi_\perp(\bp_\perp)}{|\bp_\perp|} 
\lt(\nabla_{\bp_\perp}-i{\beta}\bx_\perp\rt)\nn
  \bar W. 
\eeq
Note that the longitudinal momentum $p$ plays 
the role of a parameter in eq. (\ref{go-para}) which then
can be solved in the direction of increasing $z$ as an evolution equation with
initial data given  at a fixed $z$.  

As before we can obtain the scaling behaviors of
spatial spread, coherence length and coherence bandwidth
by dimensional analysis. 
 Let $\sigma_*$ 
the spatial spread in the transverse coordinates $\bx_\perp$, $\ell_c$ the coherence length in the transverse dimensions, $\beta_c$ the coherence bandwidth
and $L$ the distance of propagation.
We then seek the following change of
variables
\beq
\label{change2}
\tilde\bx_\perp=\frac{\bx_\perp}{\sigma_* k},\quad
\tilde\bp_\perp=\bp_\perp k\ell_c, \quad\tilde z=\frac{z}{Lk},
 \quad
\tilde\beta=\frac{\beta}{\beta_c}
\eeq
 to remove all the physical parameters from
(\ref{go-para})
\commentout{ and to aim for
the form free of physical parameters
\beq
\label{go-para-std}
\Big[\partial_{\tilde z}+{\tilde \bp_\perp\cdot\nabla_{\tilde \bx_\perp} \Big]\bar W}
&=&
\lt(\nabla_{\tilde\bp_\perp}-i{\tilde\beta}\tilde\bx_\perp\rt)\cdot |\tilde\bp_\perp|^{-1} \Pi_\perp(\tilde\bp_\perp)
\lt(\nabla_{\tilde\bp_\perp}-i{\tilde\beta}\tilde\bx_\perp\rt)
  \bar W. 
\eeq
}
Following the same line of reasoning, we obtain that
$\ell_c\sigma_*\sim L/k,\quad \sigma_*/\ell_c\sim 1/\beta_c,\quad\ell_c\sim C_\perp^{-1/3} L^{-1/3} k^{-1}
$
and hence
$
\sigma_*\sim C_\perp^{1/3} L^{4/3},\quad
\beta_c\sim C_\perp^{-2/3} L^{-5/3}k^{-1}.
$

\section{Small scale asymptotic}
On the  scale below the transport mean-free-path $\ell_{*}$  the scattering is extremely
anisotropic and the scattering amplitude is highly peaked
in the forward direction. 
This observation leads to a paraxial
approximation of 2f-RT which turns out to be
analytically solvable. 

Let $z$ be the direction of propagation of a collimated beam.
On the  scale below $\ell_{*} $
 the  2f-WD near the source point
would be highly concentrated at the longitudinal momentum,
say, 
$p=1$. Hence  we may assume that the projection $\Pi(\bp)$ in
(\ref{diff-coeff2}) is effectively just the projection onto the transverse plane coordinated by $\bx_\perp$
and we can approximate eq. (\ref{go}) by 
\beq
\label{para}
\Big[\partial_{z}+\bp_\perp\cdot\nabla_{\bx_\perp} \Big]W 
=\frac{C_\perp}{4k}\lt(\nabla_{\bp_\perp}-i{\beta}\bx_\perp\rt)^2
 W 
\eeq
where 
\beqn
C_\perp=\frac{\pi}{2}\int \Phi(0,\bq_\perp)|\bq_\perp|^2
d\bq_\perp.
\eeqn
Here we have assumed, for simplicity, the transverse isotropy of $\Phi$. 
Eq. (\ref{para}) is  another  form of paraxial approximation for which
only the one-sided (incoming) boundary condition
(at $z=$ const.)  is needed.  

We use the change of variables (\ref{change2}) with
$\ell_c\sim k^{-1}\ell_{*}^{-1/2}C_\perp^{-1/2},
\sigma_*\sim \ell_{*}^{3/2} C_\perp^{1/2},
\beta_c\sim k^{-1}C_\perp^{-1} \ell_{*}^{-2}
$
to remove the physical parameters from eq. (\ref{para}).
The transport mean-free-path $\ell_{*}$  can be determined by setting $\ell_c\sim 1$, i.e. 
$
\ell_{*}\sim k^{-2}C_\perp^{-1}. $
\commentout{to remove all the physical parameters from
(\ref{go}) and to aim for
the form
\beq
\label{fp'}
\partial_{\tilde z} W+\tilde\bp_\perp\cdot\nabla_{\tilde\bx_\perp} W
=\lt(\nabla_{\tilde\bp_\perp}-i{\tilde\beta}\tilde\bx_\perp\rt)^2W.
\eeq
The same reasoning as above now leads to
$
\ell_c \sigma_*\sim L/k,\quad \sigma_*/\ell_c \sim {1}/{\beta_c},\quad 
\ell_c\sim k^{-1}L^{-1/2}C^{-1/2}
$
and hence
$
\sigma_*\sim L^{3/2} C^{1/2},\quad
\beta_c\sim k^{-1}C^{-1} L^{-2}.
$
}
Performing the  inverse Fourier
transform in $\tilde\bp$ on the rescaled equation we obtain 
\beq
\label{mean-eq2}
\partial_{\tilde z} \Gamma
-{i}\nabla_{\tilde\by_\perp}\cdot\nabla_{\tilde\bx_\perp} \Gamma 
&=&-\big|\tilde\by_\perp+{\tilde\beta}\tilde\bx_\perp\big|^2
\Gamma\eeq
 which is the governing equation for the two-frequency coherence
 $\Gamma$. 
 By a simple change of coordinates, eq. (\ref{mean-eq2}) can be converted into a form
 similar to the time dependent Schr\"odinger equation with
 a (purely imaginary) quadratic potential and then solved analytically. Let $\Delta\br=\by_\perp+\tilde\beta\bx_\perp$ and $\Delta\br'=\by'_\perp+\tilde\beta\bx'_\perp$ be
the  field point offset   and the source point offset, respectively, measured in the unit of
central wavelength.
The  propagator  for the initial value problem  
from the source point $(\tilde\bx_\perp,\Delta\br)$ to the field point $(\bx'_\perp,\Delta\br')$ is
given by \cite{2f-rt-josa}
\beq
\nn
&&\frac{(i4\tilde\beta)^{1/2}}{(2\pi)^2\tilde z\sinh{\big[(i4\tilde\beta)^{1/2}\tilde z\big]}}  
e^{\frac{1}{i4\tilde\beta\tilde z}
\lt|\Delta\br-2\tilde\beta\tilde\bx_\perp-\Delta\br'+2\tilde\beta\bx'_\perp\rt|^2}
\\
\nn&&\times e^{{-\frac{\coth{\lt[(i4\tilde\beta)^{1/2}\tilde z\rt]}}{(i4\tilde\beta)^{1/2}}
\lt|
\Delta\br
-\frac{\Delta\br'}{
\cosh{\lt[(i4\tilde\beta)^{1/2}\tilde z\rt]}}\rt|^2}}\\
&&\times
 e^{-\frac{\tanh{\lt[(i4\tilde\beta)^{1/2}
\tilde z\rt]}}{(i4\tilde\beta)^{1/2}}
\lt|\Delta\br'\rt|^2}\label{asym}
\eeq
which converges, in the limit $\tilde\beta \downarrow 0$, to the propagator for $\tilde\beta=0$
\beq
{(2\pi \tilde z)^{-2}} e^{\frac{i}{\tilde z}
(\tilde\bx_\perp-\bx_\perp')\cdot(\Delta\br-\Delta\br')}e^{-\frac{\tilde z}{3}(|\Delta\br|^2+\Delta\br\cdot
\Delta\br'+|\Delta\br'|^2)}.\label{asym0}
\eeq
 The quadratic-in-$\Delta\br$ nature of the exponents  appearing in (\ref{asym})-(\ref{asym0})
 is the consequence of the paraxial approximation. 
 Expression (\ref{asym0}) is related to
 the asymptotic solution of the Schwarzschild-Milne equation
 in the case of very anisotropic scattering \cite{ALN}. 

In view of (\ref{plane1})-(\ref{plane3}), to get the
the correlation of two incident plane waves  we simply express (\ref{asym}) in the variables $\tilde\bx_\perp, \bx_\perp'$ and $\tilde\by_\perp,\by'_\perp$ and 
integrate it with $e^{i\Delta\hat\bk\cdot\bx'_\perp} e^{i\hat\bk\cdot\by_\perp}$. 



The functional form  of (\ref{asym}) in its dependence on
$\tilde\beta$ and $\tilde z$ is the main characteristic of the
sub-$\ell_{*}$-scale  behavior (see Fig. 1).

\section{Conclusion and discussion}
The main contribution of the present Letter is the rigorous
derivation of the  2f-RT
equation (\ref{fp}) governing 2f-WD  in disordered media and the probabilistic representation (\ref{path}). As a result, by (\ref{2time}) 
we can express the two-space-time correlation as 
\beqn
&&\lan u(t_1,\bx_1) u^*(t_2,\bx_2)\ran\\
&\sim & e^{{i\beta\bx\cdot\bp}}\int e^{i\bp\cdot\by}  e^{ik(t_{2}-t_{1})}e^{ik\theta\ep\beta(t_{1}+t_{2})/2} \fW(\bx,\bp)
 dk d\beta d\bp
\eeqn
where $\fW$ is the solution to eq. (\ref{pract}). 
The medium characteristic enters the Fokker-Planck-like eq. (\ref{fp})  only through the momentum diffusion coefficient
 (\ref{diff-coeff}). By dimensional analysis with
(\ref{fp}) and its variants we obtain
scaling behavior of spatial  spread, coherence length and coherence
bandwidth for  isotropic and  anisotropic media. We also show
that
the paraxial regime is valid for  anisotropic scattering, giving rise to two forms of paraxial 2f-RT
equations.  Finally by solving one of the paraxial  equation (\ref{para}) we obtain precise
profile of the space-frequency correlation on the scale
below the transport mean-free-path.

\begin{figure}
\centerline{\includegraphics[width=7cm, totalheight=5cm]{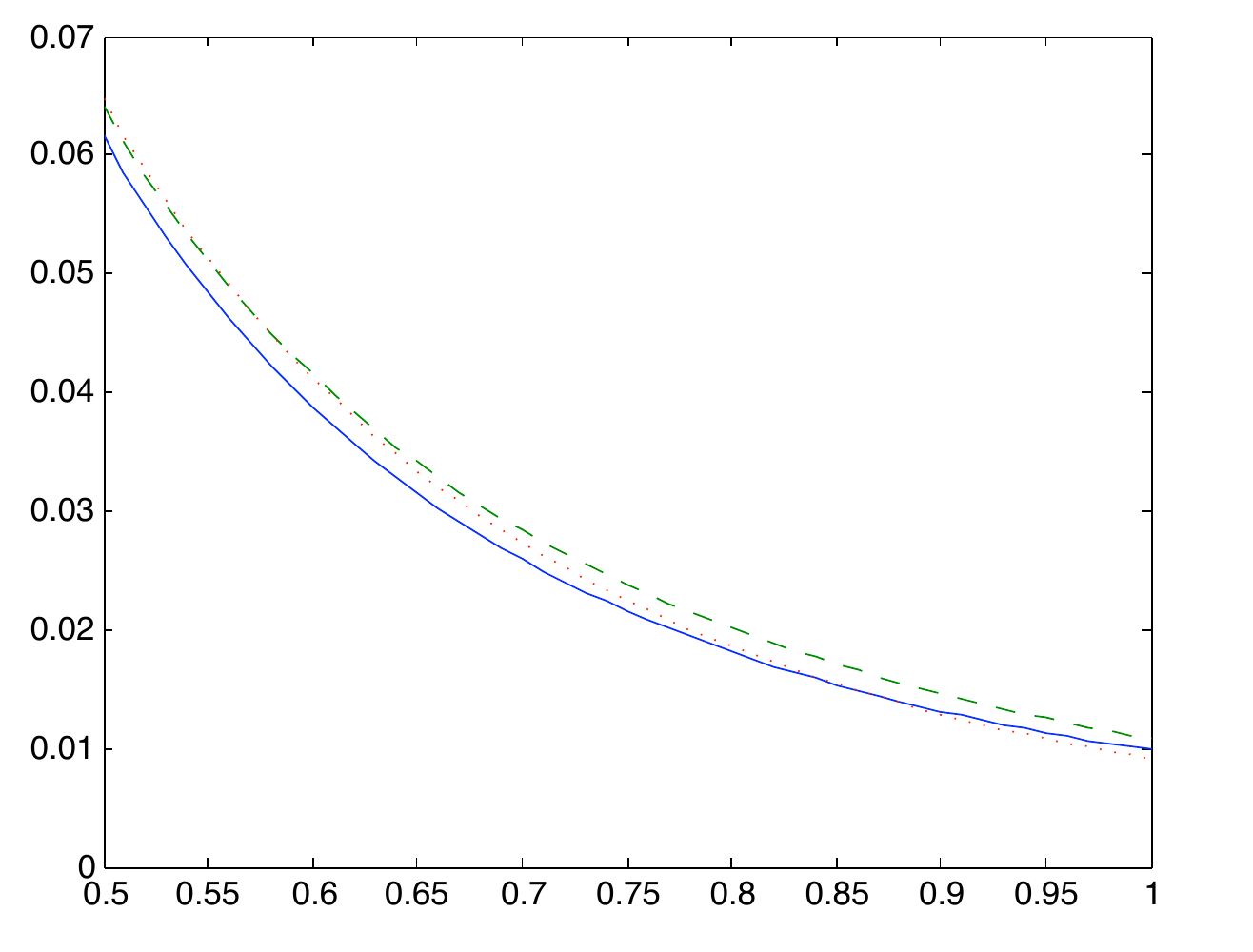}}
\caption{The absolute value of (\ref{asym}) as a function
of $\tilde z\in [0.5,1]$ for $\Delta\br=\Delta\br'=1,
\tilde\beta=0.3, 1, 3.3$ in solid, dashed and dotted lines, respectively. }
\end{figure}
Let us compare our results, especially (\ref{asym}),  with
the existing results in the literature which mostly concern
with the bulk behavior of the space-frequency correlations.

Since the bulk behavior concerns the scales   larger than  the transport 
mean-free-path 
 the existing results  are
 mostly based on
the diffusion approximation to
the displacement 
process $\bx(t)$ or the 
random-matrix method (see, e.g., \cite{RN}, \cite{FYM} and
references therein).
The diffusion regime 
represents an isotropic scattering under the condition of
equipartition of energy while
 the small-scale asymptotic (\ref{asym})
describes an extremely anisotropic scattering. 

Clearly  
the diffusion approximation is unsuitable for evaluating 
(\ref{path}) because of  the presence
of the It\^o integral with respect to the momentum process
$\bp(t)$. Therefore
to get the two-frequency coherence, the notion of 
the interference
of diffusions is invoked via diagrammatic techniques,
see the review \cite{RN}. 

In the diffusion approximation for isotropic media, 
the (dimentionless) $\bx$-diffusion coefficient $D_*$
can be derived from (\ref{fp}) with $\beta=0$
\beq
\label{diffusion-approx}
D_*= \frac{4k|\bp|^5}{3C}. 
\eeq 
The momentum-dependence of $D_*$ shows that
even in the diffusion approximation the momentum variable
is only hidden from sight.  With (\ref{diffusion-approx}) and
(\ref{scaling}) 
we can rewrite the scaling behaviors of the spatial spread, the spatial
frequency spread and the coherence bandwidth
 as 
$
\sigma_{x} \sim k^{-1}D^{-1/3}_{*},
\sigma_{p}\sim k D_{*}^{-1/3},\beta_{c}\sim D_{*}^{2/3}.
$

\commentout{
 By (\ref{diffusion-approx}) and  (\ref{recipro}) the (non-dimensionalized) coherence length and bandwidth can be written as
\beqn
\label{recipro2}
\ell_c\sim k^{-3/2}D_*^{1/2}L^{-1/2},
\quad
\beta_c\sim k^{-2}D_* L^{-2}
\eeqn
consistent with existing results. 
}

 The short-range correlation $C_1$ of wave intensities propagating  through disordered media is manifest in
the speckle pattern.  $C_1$  can be obtained   by  squaring 
the two-frequency coherence  of the wave fields \cite{PS} and the commonly accepted form is 
$
\exp{[-2\sqrt{2\tilde\beta}]}
$
which is just the large $\tilde\beta$ asymptotic of the squared factor $|\sinh{[(i4\tilde\beta)^{1/2}\tilde z]}|^{-2}$
at $\tilde z=1$
(see, e.g., \cite{Sha, FKL, Gen}).

More precisely,  the squared absolute value of
(\ref{asym}) for $\tilde z=1$ and median to large $\tilde\beta$
is approximately given by
\beq
\label{form}
\frac{4{\tilde\beta}}{(2\pi)^4}  
e^{-2\sqrt{2\tilde\beta}}
 e^{-\frac{\lt|
\Delta \br
\rt|^2}{\sqrt{2\tilde\beta}}}
 e^{-\frac{\lt|
\Delta\br'
\rt|^2}{\sqrt{2\tilde\beta}}}.
\eeq
Expression  (\ref{form}) is essentially the same as the paraxial approximation of 
 the short-range correlation  $C_1$ reviewed in
 \cite{RN}. 
 The multiplicative nature of (\ref{form})'s functional
 form in $\Delta\br$ and $\Delta\br'$ is consistent
 with the same structure in 
 the short range intensity correlation $C_1=A(\Delta k)F(\Delta \br) F(\Delta \br')$ discovered in \cite{SHG}. 
 Again,  the Gaussian form in (\ref{form}) is different from
 the form-factor $F$ in \cite{SHG} due to the paraxial approximation
 made in obtaining (\ref{form}). 

\commentout{
According to  the existing results about the bulk behavior of $C_1$,  the decay in $\Delta\br, \Delta\br'$ is exponential with a rate independent of the frequency shift
and the  propagation distance \cite{FKL}, \cite{RN}, \cite{SHG}. 
On the other hand,  (\ref{form})  interpolates 
well the bulk behavior. 
}

The long- and infinite-range correlations, represented by
$C_{2}$ and $C_{3}$ respectively, can also be obtained
by our method, \cite{RN, SHG, Sha2, SM, GSN}. The calculation  is much more involved and will be presented
elsewhere. 
\commentout{
\beq
\lt\{\begin{array}{ll}
\label{form}
\frac{(i4\tilde\beta)^{1/2}}{(2\pi)^2\tilde z}  
e^{-\tilde z(i4\tilde\beta)^{1/2}}
 e^{-\frac{\lt|
\Delta \br
\rt|^2}{(i4\tilde\beta)^{1/2}}}
 e^{-\frac{\lt|
\Delta\br'
\rt|^2}{(i4\tilde\beta)^{1/2}}},& \tilde\beta>0\label{asym2}\\
(2\pi \tilde z)^{-2} e^{-\frac{\tilde z}{3}(|\Delta\br|^2+\Delta\br\cdot
\Delta\br'+|\Delta\br'|^2)},&\tilde\beta=0.
\end{array}\rt.
\eeq
}

\commentout{
\subsection{Diffusion approximation}

The usual diffusion scaling amounts to 
\beq
\bx\to\frac{\bx}{\lambda},\quad \beta\to {\beta}{\lambda},\quad
F\to {F}{\lambda^2},\quad\lambda\to 0
\eeq
and posits the asymptotic expansion $\lan \bar W\ran=W_0+\lambda W_1+\lambda^2 W_2+\cdots.$

The leading order equation is
\beqn
\lt({i}\nabla_\bp-\frac{\beta}{k}\bx\rt)
\cdot \bD\cdot 
\lt(i\nabla_\bp-\frac{\beta}{k}\bx\rt)W_0=0
\eeqn
which can be solved by the ansatz
\beqn
W_0=\rho(\bx,|\bp|) e^{-i\bx\cdot\bp\beta/k}
\eeqn
with an, thus far,  arbitrary function $\rho$  whose
governing equation will be determined later. 

The next order equation is
\beqn
\bp\cdot\nabla_\bx W_0=-\frac{1}{4k^2}\lt({i}\nabla_\bp-\frac{\beta}{k}\bx\rt)
\cdot \bD\cdot 
\lt(i\nabla_\bp-\frac{\beta}{k}\bx\rt) W_1
\eeqn
whose solvability
condition is
\beqn
\int_{|\bp|=c}
\bp\cdot \Big[\nabla_\bx -\frac{i\beta}{k}\bp\Big]\rho(\bx,|\bp|)d\bp
=0,  \quad \forall c>0.
\eeqn
Noting the first term drops out, we see that the solvability
condition reduces to
\beqn
\int_{|\bp|=c} \rho(\bx,|\bp|)dA=0,\quad\forall c>0,
\eeqn
which is impossible to satisfy unless $\beta=0$ or $\rho=0$ identically. 
Finally the governing equation for $\rho$ is the solvability condition
for $W_2$:
\beqn
\int_{|\bp|=\hbox{const.}}  \Big[
\bp\cdot\nabla_\bx W_1-\lan F\ran\Big] e^{i\bp\cdot\bx\beta/k}dA=0.
\eeqn
}


\end{document}